\documentclass[fleqn,twoside]{article}
\usepackage{espcrc2}

\newcommand{\AmS}{{\protect\the\textfont2
  A\kern-.1667em\lower.5ex\hbox{M}\kern-.125emS}}

\newcommand{\esup}{\EE_{\rm S}}

\newcommand{\bpsi}{{\Psi}}
\newcommand{\bfi}{{\Phi}}

\newcommand{\dth}{\pa_\theta}
\newcommand{\tQ}{{\tilde Q}}
\newcommand{\hO}{{\hat\Omega}}
\newcommand{\hA}{{\hat A}}

\newcommand{\hd}{{\hat d}}
\newcommand{\hF}{{\hat F}}
\newcommand{\cas}{{\mbox{\footnotesize$\cal S$}}}

\def\ii{\'\i}

\def\ii{\'\i}

\def\ftoday{{\sl {Le \number\day \space\ifcase\month 
\or janvier\or f\'evrier\or mars\or avril\or mai
\or juin\or juillet\or ao\^ut\or septembre\or octobre
\or novembre \or d\'ecembre\fi\space \number\year}}}    
\def\ptoday{{\sl {\number\day \space de\space \ifcase\month 
\or janeiro\or fevereiro\or mar{\c c}o\or abril\or maio
\or junho\or julho\or agosto\or setembro\or outubro
\or novembro \or dezembro\fi\space de\space \number\year}}}    
\def\gtoday{{\sl {Den \number\day. \ifcase\month 
\or Januar\or Februar\or M\"arz\or April\or Mai
\or Juni\or Juli\or August\or September\or Oktober
\or November \or Dezember\fi\space \number\year}}}    
\def\today{{\sl {\ifcase\month
\or January\or February\or March\or April\or May
\or June\or July\or August\or September\or October
\or November \or December\fi \space\number\day,\space 
                                            \number\year}}}

\renewcommand{\d}{\delta}         
\newcommand{\e}{\varepsilon}

\newcommand{\m}{\mu}
\newcommand{\n}{\nu}
\newcommand{\om}{\omega}         \newcommand{\OM}{\Omega}
\newcommand{\p}{\psi}              
\newcommand{\s}{\sigma}           
\newcommand{\te}{\theta}         
\newcommand{\f}{{\phi}}           \newcommand{\F}{{\Phi}}

\newcommand{\EE}{{\cal E}}

\newcommand{\HH}{{\cal H}}

\newcommand{\OO}{{\cal O}}
\newcommand{\PP}{{\cal P}}

\newcommand{\es}{\\[1.5mm]}

\newcommand{\sla}{\raise.15ex\hbox{$/$}\kern -.57em} 
\newcommand{\Sla}{\raise.15ex\hbox{$/$}\kern -.70em}

\newcommand{\lp}{\left(}\newcommand{\rp}{\right)}

\newcommand{\complex}{{\kern .1em {\raise .47ex
\hbox {$\scriptscriptstyle |$}}
    \kern -.4em {\rm C}}}
\newcommand{\real}{{{\rm I} \kern -.19em {\rm R}}}
\newcommand{\rational}{{\kern .1em {\raise .47ex
\hbox{$\scripscriptstyle |$}}
    \kern -.35em {\rm Q}}}
\renewcommand{\natural}{{\vrule height 1.6ex width
.05em depth 0ex \kern -.35em {\rm N}}}

\newcommand{\tr}{{\rm {Tr} \,}}
\newcommand{\half}{\dfrac{1}{2}}
\newcommand{\pa}{\partial}

\newcommand{\dfrac}[2]{{\displaystyle{\frac{#1}{#2}}}}
\newcommand{\dsum}[2]{\displaystyle{\sum_{#1}^{#2}}}   
\newcommand{\dint}{\displaystyle{\int}}

\newcommand{\twiddle}{\lower.9ex\rlap{$\kern -.1em\scriptstyle\sim$}}

\newcommand{\equ}[1]{(\ref{#1})}
\newcommand{\eq}{\begin{equation}}
\newcommand{\eqn}[1]{\label{#1}\end{equation}}
\newcommand{\eea}{\end{eqnarray}}
\newcommand{\eqa}{\begin{eqnarray}}
\newcommand{\eqan}[1]{\label{#1}\end{eqnarray}}
\newcommand{\ba}{\begin{array}}
\newcommand{\ea}{\end{array}}
\newcommand{\eqac}{\begin{equation}\begin{array}{rcl}}
\newcommand{\eqacn}[1]{\end{array}\label{#1}\end{equation}}

\newcommand{\bz}{\begin{enumerate}}
\newcommand{\ez}{\end{enumerate}}

\title{Observables in Topological Theories: A Superspace Formulation}

\author{Jos\'e Lu\ii s Boldo\address[UFES]{Departamento de F\'{\i}sica, CCE,
Universidade Federal do Esp\'{\i}rito Santo (UFES), \\
Av. Fernando Ferrari, s/n, BR-29060-900 - Vit\'oria - ES (Brasil)}%
\thanks{Supported in part by the Conselho Nacional de
Desenvolvimento Cient\'\i fico e Tecnol\'ogico (CNPq -- Brazil).},
 Clisthenis P. Constantinidis\addressmark[UFES]$^*$,
 Fran\c cois Gieres\address[IPN]{Institut de Physique Nucl\'eaire,
Universit\'e Claude Bernard (Lyon 1), \\
43, boulevard du 11 novembre 1918,
      F - 69622 - Villeurbanne (France)}, 
 Matthieu Lefran\c cois\addressmark[IPN]
\\ and Olivier Piguet\addressmark[UFES]$^*$%
}

\begin{document}

\begin{abstract}
Observables of topological Yang-Mills theory were defined by Witten
as the classes of an equivariant cohomology. We propose 
to define them alternatively as the
BRST cohomology classes of a superspace version of the theory,
where BRST invariance is associated to super Yang-Mills invariance.
We provide and discuss the general solution of this cohomology.$^1$
\vspace{1pc}

\end{abstract}
\maketitle

\section{Introduction}

\footnotetext{Prepared for International Conference on Renormalization Group 
and Anomalies in Gravity and Cosmology (IRGA 2003), Ouro Preto, MG, Brazil, 17-23 Mar 2003. 
Published in Nucl.Phys.Proc.Suppl.127\,(2004)\,30.}

Observables in topological theories are global, such as knot invariants in
Chern-Simons theory, Wilson loops, etc. The problem of finding them all
is a problem of ``equivariant cohomology'', as already pointed out 
by Witten in 1988 in the framework of 4-dimensional topological 
Yang-Mills theory\cite{witten,ouvry,delduc}. 
Soon after, a superspace description of the latter model and its 
symmetries has been introduced~\cite{horne,carvalho,thompson}.

Topological Yang-Mills theory is characterized by a local symmetry
generator $\tQ$ which is nilpotent in a functional space constrained by
gauge invariance, and equivariant cohomology is the cohomology of $\tQ$
restricted to this space. On the other hand, in the superspace approach
which we shall use hereafter, the problem is transformed to that of 
looking for Yang-Mills BRST cohomology in a supersymmetric context where
$\tQ$ corresponds to a strictly nilpotent supersymmetry generator $Q$.

Whereas the complete solution of equivariant cohomology 
is rather difficult to
obtain~\cite{witten,delduc}, we hope that our proposal
of working with a super-Yang-Mills cohomology should simplify the task,
and also make it more systematic and manageable for possible
applications to more general models, such as topological gravity and
models with more than one supersymmetry generator~\cite{inprogress}.

An extended paper on this subject is available~\cite{bcglp}.

\section{Topological Yang-Mills Theory in Four Dimensions}\label{sect2}

The purpose of four-dimensional topological Yang-Mills theory is to 
describe instanton or anti-instanton gauge field configurations. Denoting
by $a= a_\m^a(x) T_a dx^\m$ the Yang-Mills gauge connection, 
where the matrices $T_a$ form a basis 
of the gauge Lie algebra, with
\eq
[T_a,T_b] = f_{ab}{}^c T_c\ ,\quad \tr (T_aT_b)= \d_{ab} \ ,
\eqn{Lie-alg}
we may write the instanton equation as the self-duality condition
\eq
f_{\m\n}(x) = \half \e_{\m\n\rho\s}f^{\rho\s}(x)\ ,
\eqn{self-dual}
where $f_{\m\n}$ is the Yang-Mills field strength associated to the 
connection $a$, defined by
\eq
f(a) = da+a^2 = \half
f_{\m\n} 
(x) dx^\m dx^\n\ .
\eqn{curv}
Note that the self-duality condition 
\equ{self-dual}
is obviously invariant 
under the usual (infinitesimal) Yang-Mills (YM) gauge transformation
\eq
\d_{(\om)} a = d\om+[a,\om] \ . 
\eqn{g-transf-a}
Witten's way~\cite{witten} of generating these instanton configurations
 is to consider
equation
\equ{self-dual} as a gauge fixing condition for a {\it local shift
symmetry} defined infinitesimally by its infinitesimal generator 
$\tQ$:
\eq
\tQ a(x) = \p(x)\ .
\eqn{shift-susy-a}
The coefficients $\p_\m(x)$ of the infinitesimal parameter 
1-form $\p$ are taken as
anticommuting 
fields, i.e. as ghost fields, the generator $\tQ$ 
being thus interpreted as the BRST
operator associated with the local shift symmetry. However,
the transformation
\equ{shift-susy-a} is gauging away all 4 degrees of freedom of $a$, 
whereas there are only 3 independent equations in the condition 
\equ{self-dual}. Therefore 
one has to gauge away one of the four degrees of freedom of $\p$. 
This is done by
imposing invariance under a local transformation of the ghost $\p$:
\eq
\tQ \p = -D_a\f \equiv -d\f -[a,\f]\ ,\quad \tQ \f =0\ ,
\eqn{shift-susy-psi}
where the ghost for ghost $\f(x)$ is a 
commuting
0-form. The 
$\tQ$-invariance of $\f$
is needed for the sake of the nilpotency of the BRST
operator $\tQ$, which however holds only up to a field dependent
YM-gauge transformation generalizing \equ{g-transf-a}, defined by
\eq\ba{l}
\d_{(\om)} a = d\om+[a,\om]\ ,\quad
\d_{(\om)}\p= [\p,\om]\ ,\es \d_{(\om)}\f=[\f,\om] \ ,
\ea\eqn{g-transf}
but
with the infinitesimal parameter $\om$ replaced by the ghost for 
ghost field $\f$. One checks indeed that
\eq
\tQ^2 = \d_{(\f)}\ .
\eqn{equiv-nilp}
However $\tQ$ is 
nilpotent when applied to YM-gauge invariant 
quantities. This led Witten
to define the {\it observables} of the theory as the cohomology classes of
$\tQ$ in the space of the gauge invariant operators, which is a problem of
{\it equivariant cohomology}~\cite{ouvry,delduc}: an 
operator $\OO$ belongs to the equivariant cohomology of $\tQ$ iff
\[
\tQ \OO=0\ ,\qquad  \mbox{but}\quad \OO \not=\tQ\PP\ , 
\]
with the conditions:
\[
   \d_{(\om)}\OO=0\ ,\quad \d_{(\om)}\PP=0\ .
\]
To write an invariant action needs the introduction of other fields 
than $a, \p , \f$
which 
play the roles of antighosts and Lagrange multipliers for the self-duality 
constraint \equ{self-dual}. This aspect of the theory will not be touched 
in the 
present contribution, the reader may consult the original 
literature~\cite{witten,ouvry}. We shall concentrate on 
the ``geometrical sector''
spanned by the connection $a$ and the ghosts $\p$ and $\f$ -- which 
is what is needed for the determination of the observables.

\section{Superspace and Superforms}\label{sect3}

The shift symmetry BRST operator $\tQ$ is odd, i.e. it transforms
fermions into bosons and vice-versa, hence it has the character of a
supersymmetry (SUSY) generator. This has suggested,
already a long time 
ago~\cite{horne,carvalho,thompson}, a superspace and
superfield description. Beyond a natural supersymmetry 
operation, whose generator is denoted by $Q$, the superspace 
formalism involves 
a superconnection and a supergauge invariance. 
The supergauge invariance
may be partially fixed {\it \`a la Wess-Zumino} by setting to zero certain components
of the superconnection, leaving us with the field content described in the last
section. The shift symmetry operator $\tQ$
defined by  
(\ref{shift-susy-a},\ref{shift-susy-psi})
is then recovered as the
composition of the supersymmetry operation $Q$ with a particular 
field dependent
supergauge transformation~\cite{horne,carvalho,thompson,bcglp}.

Superspace is introduced by
enlarging the $d$-dimensional spacetime, of coordinates $x^\m\;
(\m=0,\cdots,d-1)$, with the addition of
one fermionic dimension\footnote{This construction 
may be generalized to the case of $N$ SUSY generators, in a
superspace with $N$ fermionic dimensions~\cite{thompson}.}
described by a 
Grassmann (i.e. anticommuting) coordinate $\theta$.
One then defines a {\it superfield} as a superspace function 
\[
S(x,\theta) = s(x) + \theta s'(x)\ ,
\]
linear in $\theta$ since $\theta^2=0$,
which by definition transforms under 
an infinitesimal 
SUSY transformation as 
\eq
QS=\dth S\ ,
\eqn{SUSY-tr}
\eq
\mbox{or, in components:}\quad Qs=s'\ ,\quad Qs'=0\ .
\eqn{SUSY-tr-comp}
By construction the SUSY operator is strictly 
nilpotent:
$Q^2=0$.
Introducing the differential $d\theta$ -- a commuting quantity -- 
we may define {\it $p$-superforms},
\[
\hO_p (x,\theta) = \dsum{k=0}{p} \, \OM_{p-k}(x,\theta)(d\theta)^k \ ,
\]
where $\OM_{p-k}(x,\theta) = \om_{p-k}(x) + \theta\om_{p-k}'(x)$ is a
$(p-k)$-form in $x$-space with superfield coefficients.
The {\it superspace exterior derivative} reads
\eq
\hd = d+d\theta\dth 
= dx^\m\pa_\m+d\theta\dth\ ,\ \ \mbox{with}\ \hd^2=0\ .
\eqn{s-der}
A supergauge theory will be based on a {\it superconnection}, the
1-superform\footnote{All forms and superforms in this paper 
are Lie algebra valued: 
$\F=\F^aT_a$, see \equ{Lie-alg}.}
\[\ba{l}
\hA = A(x,\theta) + d\theta  \, A_\theta(x,\theta)\es
\phantom{\hA }
=  a(x)+\theta\p(x) +  d\theta\lp\chi(x)+\theta\f(x)\rp\ ,
\ea\]
a  Faddeev-Popov ghost 0-superform
\[
C (x,\theta)  =  c(x) + \theta  c'(x)\ ,
\]
and the corresponding {\it BRST transformations}
\eq
\cas \hA = -( \hd C +[ \hA ,C ] )\ ,\ 
\cas C =  -C^2 \ (\cas^2=0)\ .
\eqn{BRST}
The latter read, in components:
\eq\ba{ll}
\cas a =  -Dc  \ ,\ &\cas\p = - [c,\p]  -Dc' \ ,\\
\cas\chi = - [c,\chi] - c'\, ,\ &\cas\f = - [c,\f] - [\chi,c'] \ ,\\ 
\cas c  =  -c^2\ , \ &  \cas c' =  -[c,c'] \ .
\ea\eqn{BRST-comp}


%
The {\it Wess-Zumino gauge} mentioned at the beginning of this section
consists in putting $\chi=0$. The $\tQ$-transformations of 
Section \ref{sect2} are then reproduced by adding to the 
$Q$-transformations \equ{SUSY-tr} a particular
supergauge transformation, namely a BRST transformation 
\equ{BRST-comp} with $c=0$ and $c'=\f$. 
This shows the equivalence of the superspace 
approach with the formulation originally 
proposed by Witten, the latter being a Wess-Zumino gauge fixed version
of the former.
We shall however keep to the
superspace formalism for the rest of this paper, only specializing to 
the Wess-Zumino gauge for comparisons with the literature.

\section{Observables in the Superspace Formalism}\label{sect4}

By contrast to 
the equivariant cohomology 
defined by 
$\tQ$, 
the (unrestricted) cohomology of 
 $Q$ is trivial: every form or superform which is 
$Q$-invariant is the $Q$-variation of
another form or superform. This follows~\cite{book,barnich} from 
the fact that all fields are 
grouped in doublets $\{s(x),\,s'(x)\}$ as in \equ{SUSY-tr-comp}.
However, 
the cohomology that is {\it a priori} not  
trivial is indeed the cohomology of the
BRST operator (\ref{BRST},\ref{BRST-comp}). This observation 
suggests to define an observable $\OO_{(d)}$ as
a BRST cohomology class:
\eq
\cas \OO_{(d)} = 0\ ,\quad\mbox{but}\quad \OO_{(d)}\not= \cas \PP_{(d)}\ ,
\eqn{s-cohom}
where $\OO_{(d)}$ and $\PP_{(d)}$ are both $d$-dimensional\footnote{The space-time
dimension will not be fixed a priori.
The space-time integrals are performed over 
an arbitrary $d$-manifold $M_d$.}
space-time integrals:
\[
\OO_{(d)}=\dint_{M_d}\om_d(x)\ ,\quad \PP_{(d)}=\dint_{M_d}\f_d(x)\ ,
\]
with the SUSY conditions:
\eq
   Q\OO_{(d)}=0\ ,\quad Q\PP_{(d)}=0\ .
\eqn{SUSY-cond}
The latter condition and 
the triviality of the cohomology of $Q$ implies now that, up to a possible
total derivative which can be discarded without loss of generality, the
integrand of $\OO_{(d)}$ -- the $d$-form $\om_d$ -- is $Q$-exact:
\[
\om_d(x) = Q\OM_d(x,\theta)\ ,
\]
hence
\eq
\OO_{(d)} = \dint_{M_d}Q\OM_d(x,\theta) = \dint_{M_d}\dth\OM_d(x,\theta)\ ,
\eqn{s-int}
which is the superspace integral of a {\it superfield form}, i.e. of a
space-time form whose coefficients are superfields. From now
on we shall  work with such superfield forms.

\section{General Solution}\label{sect5}

We want to find the 
general solution of the BRST cohomology problem \equ{s-cohom} with the 
SUSY inv\-ar\-ian\-ce condition \equ{SUSY-cond}. The latter condition
being automatically fulfilled using superspace integrals such as 
\equ{s-int}, we are left with the BRST cohomology equation 
\equ{s-cohom}. It implies for the integrand $\OM_d$ $\equiv$ $\;^{D-d}\OM_d^0$  
in \equ{s-int} the condition
\eq
\cas \;^{D-d}\OM_{d}^{0} + d \;^{D-d}\OM_{d-1}^{1} 
+ Q \;^{D-d-1}\OM_{d}^{1} = 0 \ ,
\eqn{des-eq-0}
where $\;^{D-d}\OM_{d-1}^{1}$ and $\;^{D-d-1}\OM_{d}^{1}$ are some superfield
forms. The indices $p$, $g$ and $s$ of $\;^s\OM_p^g$ mean,
respectively, its space-time form degree, ghost number and SUSY number.
The ghost number is equal to the 
degree in the superfield ghost $C$ or its components
$c$, $c'$. 
The SUSY number, which is indeed the ``ghost number'' corresponding to
local shift symmetry, is defined by giving the value $s=1$ to the SUSY
operator $Q$ (hence $s=-1$ for the $\te$ coordinate) and $s=0$ for
the Yang-Mills connection $a(x)$. All the forms 
$^{\cdot}\OM_{\cdot}^{\cdot}$ are taken as 
polynomials in the superfields $A$, $A_\te$, $C$ and their 
derivatives $d\;(\cdots)$, $\dth\;(\cdots)$, which defines a
functional space denoted by $\esup$.

Now, nilpotency of $\cas$ together
with the cohomological triviality of $d$ and $Q$ in the space $\esup$
imply that \equ{des-eq-0} is but the first one of a set of 
{\it bi-descent equations}~\cite{bcglp}:
\eq\ba{l}
\cas\;^{D-p-g}  \OM_{p}^{g} +  d\;^{D-p-g}  \OM_{p-1}^{g+1}\es
+ Q\;^{D-p-g-1}  \OM_{p}^{g+1} = 0 \es
(\,p= 0,\dots,d\;;\ g=d-p,\dots,D-p \, )   \ .
\ea\eqn{bi-descent}
The observables of dimension $d$ and SUSY number $s$ will thus 
be given
by \equ{s-int} as 
the general solution 
$\;^{s-1}\OM_d^0$ of \equ{bi-descent}, at ghost number 0, 
i.e. with $D=d+s-1$.

The general solution of the bi-descent equations
\equ{bi-descent} is divided in 
the two classes described in the two following subsections~\cite{bcglp}.

\subsection{Equivariantly trivial solutions}\label{trivial} 
The first class of nontrivial solutions of the bi-descent equations
correspond to nontrivial solutions of the BRST cohomology 
problem \equ{s-cohom} 
which are {\it trivial in the sense of the equivariant cohomology}.
They have the general form~\cite{bcglp}
\eq\ba{l}
\;^{D-d+1}\OO_{(d)}  \es =
\int_{M_d} Q \;^{D-d}\HH_d(F_A,\bpsi,\bfi,{D_A}\bpsi,{D_A}\bfi)\ ,
\ea\eqn{triv-obs}
where $\HH_d$ is a gauge invariant function of 
the superfield forms $F_A$, $\bpsi$ and $\bfi$ and their covariant
derivatives $D_A\bpsi$ = $d\bpsi+[A,\bpsi]$ and  $D_A\bfi$.
These superfield forms are the components of the supercurvature
$\hF \equiv \hd \hA + \hA^2$,  
\eq
\hF = {F_A} + \bpsi \, d\theta  + \bfi \, (d\theta)^2 \ , 
\eqn{s-curvature}
with 
\[\ba{l}
{F_A} = dA+A^2 \ , \es
\bpsi = \partial_{\theta} A + D_A A_\theta 
= \psi +  D_a\chi + O(\theta)\ ,\es
\bfi  = \partial_{\theta} A_\theta + A_\theta ^2 
= \f+ \chi^2 + O(\theta)\ .
\ea\]
Their BRST transformations read as 
\[\ba{l}
\cas F_A =  - [ C,F_A]  \ ,\quad 
\cas\bpsi = - [C,\bpsi] \ ,\es
\cas\bfi = - [C,\bfi] \ .
\ea\] 
The solutions \equ{triv-obs}, although 
nontrivial in the sense of the
BRST cohomology, are trivial in the sense of the equivariant
cohomology. 
An easy way to see this is to go to the Wess-Zumino gauge, 
where they conserve the same form as in
\equ{triv-obs}, but with $A$, $F_A$, $\bpsi$ and $\bfi$ replaced by
$a$, $F_a=da+a^2$, $\p$ and $\f$, and the operator $Q$ replaced by
$\tQ$.
The result is then explicitly given by the $\tQ$-variation of a gauge
invariant integral.

\subsection{Equivariantly nontrivial solutions}\label{non-trivial}

The second class of nontrivial solutions of the bi-descent equations 
\equ{bi-descent} 
corresponds to nontrivial solutions of the BRST cohomology 
problem \equ{s-cohom} 
which are also {\it nontrivial in the sense of the equivariant cohomology}.
They are
given in terms of superforms by the following superspace 
algorithm~\cite{bcglp}.
\vspace{3mm}

{\bf1.} Consider all the superforms $\hO_D(x,\theta)$ 
(of ghost number 0 and superform degree $D$) which are
nontrivial elements 
of the cohomology $H(\cas|\hd)$
of $\cas$ modulo $\hd$ in the space of the superforms 
consisting of polynomials 
of the basic superforms $\hA(x,\theta)$, $C(x,\theta)$, 
$\hd\hA(x,\theta)$ and $\hd C(x,\theta)$. Nontriviality in the sense of
the cohomology $H(\cas|\hd)$ for a superform $\hO$ means
\[\ba{l}
\cas\hO = 0\quad(\mbox{modulo\ }\hd)\ ,\es
\mbox{but}\quad
\hO\not= \cas{\hat\F}\quad(\mbox{modulo\ }\hd)\ .
\ea\]
These $\hO_D$ are nontrivial solutions of sets of ``superdescent equations''
\eq\ba{l}
\cas \hO_D + \hd\hO_{D-1}^1=0\, ,\
\cas \hO_{D-1}^1 + \hd\hO_{D-2}^2=0\, ,\es 
\quad\cdots \ ,\quad
\cas \hO_{0}^{D}=0 \ .
\ea\eqn{s-descent}


\vspace{3mm}

{\bf 2.} Expand $Q\hO_D = \dth\hO_D$ according to the space-time 
form degree $p$:
\eq
Q\hO_D = \dsum{p=0}{D} w_p(x) (d\theta)^{D-p} \ .
\eqn{gen-observ}
The space-time forms $w_p$ are our solutions. Indeed, 
\[
\cas w_p(x) = 0 \ (\mbox{modulo\ }d) \quad \mbox{and}\quad
Qw_p (x) = 0 \ ,
\]
which follows from applying the operator $Q$ to the first of the
superdescent equations \equ{s-descent}, and using the identities
$Q\,\hd=-\hd\,Q= -d\,Q$,
which are direct consequences 
of the definition \equ{s-der}.

We thus obtain, for a given maximum degree $D$, 
a set of observables
\eq
\OO_{(p)} = \dint_{M_p} w_p(x) \quad(p=0,\dots D)\ ,
\eqn{observables}
where the space-time forms 
$w_p$ are the coefficients of the superform
$Q\hO_D$, with $\hO_D$ representing a nontrivial 
solution of the superdescent equations \equ{s-descent}.

\subsection{Solution of the superdescent equation}

The nontrivial observables are thus completely determined from the
general solution of the superdescent equations \equ{s-descent}. The
latter is given by the generalization to the present superspace
formalism of standard results of BRST cohomology~\cite{barnich}. The
result is:
\eq\ba{l}
\hO_{D} = \theta^{\rm CS}_{r_1}(\hA) 
f_{r_2}(\hF)\cdots f_{r_L}(\hF)\ , \es
\mbox{with}\quad  D= 2 \, \dsum{i=1}{L} \, m_{r_i} -1 
\ , \quad L \geq 1 \ ,
\ea\eqn{3.xx}
where $f_r(\hF)$ is the supercurvature invariant of degree $m_r$ in 
$\hF$ corresponding to the gauge group Casimir operator of degree
$m_r$, and $\theta^{\rm CS}_{r}(\hA)$ is the associated Chern-Simons
form: 
\eq
\hd \theta^{\rm CS}_{r}(\hA) = f_r(\hF)\ .
\eqn{s-CS}
We note that the superform degree of the solution \equ{3.xx} is odd.

An equivalent and convenient way to write down the integrants
$w_p(x)$ of the observables
is given by the following
expansion of the super exterior derivative of the superform 
\equ{3.xx}:
\eq\ba{l}
\hd
\hO _D 
= f_{r_1}(\hF)  \cdots  f_{r_L}(\hF) \es
\phantom{\hd\hO ^0_D}
= f_{r_1}(F_A) \cdots f_{r_L}(F_A) \es
\phantom{\hd\hO ^0_D} \quad
+ \dsum{p=0}{D}  \;^{D+1-p}W^0_{p}  \; (d\theta)^{D+1-p}  \ ,
\ea\eqn{exp-hd-OM}
the first equality following from \equ{s-CS}.
It is indeed easy to check that the forms
\eq
w'_p(x) = \left. \;^{D-p+1}W_p^0\right|_{\theta=0}
\eqn{altern}
differ from the $w_p(x)$ by derivative terms only,
and thus may be substituted to the latter 
in the integrals \equ{observables} defining the observables.

One finally checks that, upon 
reducing the results to the Wess-Zumino gauge, one
recovers the observables originally given by Witten \cite{witten}.

\section{Example}\label{example}

We consider the case of maximum degree $D=3$. The superdescent equations
read as
\[\ba{ll}
\cas \hO_3 + \hd\hO_2^1=0\ ,\quad
&\cas \hO_2^1 + \hd\hO_1^2=0\ ,\es
\cas \hO_1^2 + \hd\hO_0^3=0\ ,\quad
&\cas \hO_0^3=0\ .
\ea\]
The unique nontrivial solution is 
\[\ba{ll}
\hO_3 =  \tr ( \hA \hd\hA + \frac{2}{3} \hA^3 ) \ ,\quad
&\hO_2^1 =  \tr (\hA \hd C ) \ ,\es
\hO_1^2 =  \tr ( C \hd C ) \ ,\quad
&\hO_0^3 =  -\frac{1}{3} \; \tr C^3 \ .
\ea\]
Note that $\hO_3$ is the Chern-Simons superform associated 
to the quadratic Casimir operator of the gauge group.
%
The observables are then given by the expansion
\eq\ba{l}
\left.\hd \hO_3^0 \right|_{\theta=0} 
= 
\left. \tr \hF ^2  \right|_{\theta=0} \es
\phantom{\left.\hd \hO_3^0 \right|_{\theta=0}}
= \tr {F_a} ^2 + 
\dsum{p=0}{3} w'_p \; 
 (d\theta)^{4-p} 
\ ,
\ea\eqn{hat-om-3-0}
with
\eq\ba{l}
w'_0 = \tr(\f^2 + 2\f\chi^2 ) \\
w'_1 =  2\tr ( \psi\f + \p\chi^2 + 
\f D_a\chi ) +d(\cdots)\ , \\
w'_2 = \tr(\p^2 + 2\f F_a + 2\p D_a \chi) 
 +d(\cdots) \\
w'_3 = 2\tr( \p F_a )  +d(\cdots) \ .
\ea\eqn{list-fi2}
 The observables are the integrals of these forms 
 (and of $\tr {F_a} ^2$) 
 on closed submanifolds of appropriate dimension. 

In the Wess-Zumino gauge:
\[\ba{l}
w'_0     =   \tr ( \f^2)  \ ,\\
w'_1      =  2\tr ( \p\f  ) +d(\cdots)\ ,\\
w'_2      = 
\tr ( 2
\f F_a + \p^2) +d(\cdots)\ ,\\
w'_3      =  2\tr (\p F_a ) +d(\cdots)
\ea\]
which corresponds to Witten's result up to total derivatives.

\section{Conclusion}

The superspace
BRST cohomology which we have proposed  as an alternative  definition 
of the observables of 
a topological theory of the Yang-Mills type, reproduces 
Witten's original result
using a rather straightforward extension 
to superspace of standard results on BRST cohomology \cite{barnich}.
We have also seen that this cohomology produces other solutions,
which are easily
proved to be trivial in the sense of equivariant cohomology. 
The main achievement has been to show \cite{bcglp}
that there are no other solutions 
to the problem of the superspace
BRST cohomology. These results lead us to wonder about the applicability 
of such an approach to
the construction of observables in more 
complex systems, for example
topological gravity and Yang-Mills theories 
with more than one supersymmetry generator. These are problems
under current investigation, and we hope to 
provide answers to these 
questions soon \cite{inprogress}.

\section*{Acknowledgments}

J.L.B., C.P.C. and O.P. thank the Con\-se\-lho Na\-cio\-nal de
Desenvolvimento Cient\'\i fico e Tecno-l\'o-gi-co (CNPq -- Brazil) for
financial support.\\
C.P.C. also thanks the Coordena\c c\~ao de Aperfei\c coamento de 
Pessoal de N\'\i vel Superior (CA\-PES -- Bra\-zil) for financial support,
and the Abdus Salam International Center for Theoretical Physics
(ICTP -- Italy) for hospitality during a three months stay
under its Associate Program.\\
O.P. acknowledges
the Institut de Physique Nuc\-l\'e\-ai\-re,
Universit\'e Claude Bernard (Lyon 1) for its 
kind hospitality during a
one month's stay as Professeur Invit\'e.


\end{document}